\documentclass[9pt]{article}
\usepackage{waspaa23,amsmath,graphicx,url,times}
\usepackage{color}
\usepackage{amssymb}
\usepackage{booktabs}
\usepackage{subfig}
\usepackage{cite}
\usepackage[font=small,skip=3pt]{caption}
\title{SEFGAN: Harvesting the Power of Normalizing Flows and GANs for Efficient High-Quality Speech Enhancement}

 \name{Martin Strauss$^{1}$\sthanks{Corresponding author: martin.strauss@audiolabs-erlangen.de},
       Nicola Pia$^{2}$,
       Nagashree K. S. Rao$^{2}$,
       Bernd Edler$^{1}$}
 \address{$^1$ International Audio Laboratories Erlangen\sthanks{A joint institution of the Friedrich-Alexander-Universität at Erlangen-Nürnberg (FAU) and Fraunhofer IIS.}, Am Wolfsmantel 33, 91058 Erlangen, Germany\\ 
          $^2$ Fraunhofer IIS, Erlangen, Germany\\
 }

\begin{document}

\ninept
\maketitle

\begin{sloppy}

\begin{abstract}
This paper proposes SEFGAN, a Deep Neural Network (DNN) combining maximum likelihood training and Generative Adversarial Networks (GANs) for efficient speech enhancement (SE). For this, a DNN is trained to synthesize the enhanced speech conditioned on noisy speech using a Normalizing Flow (NF) as generator in a GAN framework. While the combination of likelihood models and GANs is not trivial, SEFGAN demonstrates that a hybrid adversarial and maximum likelihood training approach enables the model to maintain high quality audio generation and log-likelihood estimation. Our experiments indicate that this approach strongly outperforms the baseline NF-based model without introducing additional complexity to the enhancement network. A comparison using computational metrics and a listening experiment reveals that SEFGAN is competitive with other state-of-the-art models. 

\end{abstract}


\section{Introduction}
\label{sec:intro}

The goal of Speech Enhancement (SE) is to obtain a target speech utterance from a noisy sample degraded by background noise. 
It is a widely studied subject with real world applications including front ends for speech recognition systems \cite{koizumi22b_interspeech}, hearing aids \cite{schroeter_2022} or broadcasting \cite{dnn_ds}. Typically, modern architectures are based on Deep Neural Networks (DNN) that learn to minimize a distance metric as in Conv-TasNet \cite{Luo2019ConvTasNet}. 

Deep generative models are receiving increasing popularity in recent years \cite{strauss2021, lu2022conditional, strauss_slt, richter2023}. 
Their aim is to estimate a probability distribution of clean speech samples by learning the fundamental structure of the input data.
The most prominent examples include Generative Adversarial Networks (GANs) \cite{fu21_interspeech, cao22_interspeech} or Denoising Diffusion Probabilistic Models (DDPMs) \cite{lu2022conditional, richter2023, serrà2022universal}. 

GANs in particular have repeatedly shown to be an effective tool to achieve high denoising performance with approaches like MetricGAN+ \cite{fu21_interspeech} and CMGAN \cite{cao22_interspeech} reporting strong evaluation results. Still, these results are at times hard to reproduce or retrain due to the unstable nature of GAN training. 

Recently, there has been growing interest in diffusion probabilistic models. Lu et al.~\cite{lu2022conditional} were the first to apply conditional diffusion techniques to time-domain SE. 
In the following, Richter et al.~\cite{richter2023} built on previous work \cite{welker22speech} demonstrating strong performance on SE and dereverberation in the complex short-time Fourier transform (STFT) domain. However, DDPMs require high computational resources at inference time \cite{song2021denoising}.\\ \indent
Normalizing flows (NF) are another family of generative models based on invertible differentiable transformations. They allow for direct density estimation via maximum likelihood optimization, offer high-quality generation, and allow for highly parallelizable inference. 
In \cite{strauss2021} a time domain NF approach inspired by WaveGlow~\cite{Prenger2019} was proposed for SE.
This approach was extended in \cite{strauss_slt} with an all-pole gammatone filterbank as conditional pre-processing. NFs have strong theoretical properties, but current NF-based SE systems still show lower performance compared to DDPMs and GANs.\\ \indent
Although DDPMs and pure GANs show strong evaluation results, they are difficult to train and require lots of engineering for good performance and fast inference.
Moreover, it is often difficult to improve the performance of generative models by using specific domain knowledge like a perceptually motivated filterbank.
Hence, there is a clear need for solutions that are easy to train, but still offer high performance at acceptable complexity. \\ \indent
In this work, we propose \textbf{SEFGAN} (\textbf{S}peech \textbf{E}nhancement \textbf{F}low \textbf{GAN}), a combination of a generative NF-based model and a GAN framework. Therefore, an existing NF system for SE is modified by an additional conditional network, and is then used as a GAN generator. 
Our experiments indicate that the presented approach strongly improves the performance of a pure NF or pure GAN training. \\ \indent
In addition, we notice that by simply training a pretrained NF model adversarially we can obtain good speech quality, but the capability of the model to perform log-likelihood estimation is completely lost.
For this reason, similar to \cite{flowgan} we apply a hybrid approach, labeled \textbf{SEFGAN-hybrid}, which is simultaneously trained to maximize log-likelihood and minimize the adversarial loss.
This not only allows the model to achieve high-quality SE while maintaining its capability to estimate the log-likelihood for a given input, but is a particularly stable training procedure.\\
\indent SEFGAN outperforms state-of-the-art systems according to  computational metrics. Our listening test shows that the quality of the denoised audio is close to one of the most recent diffusion-based models.
Moreover, it offers significantly faster inference than diffusion models, utilizes less parameters and offers reliable log-likelihood estimation.
 
	

\section{SEFGAN}
\label{sec:sefgan}
\subsection{Normalizing Flow-based speech enhancement}
A NF is a bijective mapping of two random variables $\textbf{x}, \textbf{z} \in \mathbb{R}^{D}$ via a differentiable function $\textbf{z} = f(\textbf{x})$ with differentiable inverse. While $\textbf{z}$ is commonly assumed to be sampled from a standard Gaussian, the variable $\textbf{x}$ can be derived from a distribution of arbitrary complexity.
In this set up the change of variables formula $p_x(\textbf{x}) = p_z(\textbf{z}) \left| \partial \textbf{z}/ \partial \textbf{x}\right|$ 
permits to train the NF via explicit log-likelihood maximization.

Let's now define a noisy observation $\textbf{y}\in \mathbb{R}^{N}$ as the combination of a clean speech utterance $\textbf{x}\in \mathbb{R}^{N}$ and a disturbing background noise $\textbf{n}\in \mathbb{R}^{N}$, i.e. $\textbf{y} =  \textbf{x} + \textbf{n}$. All components are in time domain, single channel and with $N$ samples per utterance.

Further, let's define a random variable  $\textbf{z}\in \mathbb{R}^{N}$ with $\textbf{z}$ being sampled from a standard Gaussian distribution. The goal of SE is to recover $\textbf{x}$ from $\textbf{y}$. 
In NF-based SE this is achieved by learning a bijective mapping conditioned on the noisy utterance $\textbf{y}$ via minimizing the negative log-likelihood (NLL) of the target distribution $p_x(\textbf{x}|\textbf{y}; \theta)$, i.e., 
\begin{equation}
	\mathcal{L}_{nll} = -\log p_z(\textbf{z}|\textbf{y}) - \log \left\lvert \det \left( J(\textbf{x}, \textbf{y}) \right) \right\rvert,
\end{equation}
\noindent where $\textbf{z}=f_{\theta}(\textbf{x}, \textbf{y})$ with $\theta$ as the model parameters and $J(\textbf{x}, \textbf{y})=\partial \textbf{z}/ \partial \textbf{x}$ represents the Jacobian.
At inference time, a random sample $\textbf{z} \sim p_z(\textbf{z})$ and a noisy representation $\textbf{y}$ are used to generate the enhanced speech sample via the inverted flow network $f_{\theta}^{-1}$, mapping from $\textbf{z}$ to $\hat{\textbf{x}} \approx \textbf{x}$. 

\subsection{The baseline architecture}
\label{sec:proposed}
\noindent The baseline architecture is derived from the SE-Flow model used in \cite{strauss2021} which is based on WaveGlow \cite{Prenger2019}.
It consists of a sequence of flow blocks, with each block containing a 1$\times$1 invertible convolution and an affine coupling layer for transformation. 
The network processes multiple channels of the input at once with the input signal being subsampled by a factor $s$ to create a multichannel input $x \in \mathbb{R}^{s \times (N/s)}$. In each coupling layer, the input to the layer is split into two halves, where one half is used as input to a subnetwork $f_{sub}$ estimating $s$ and $t$ as the affine scaling and translation parameters. Here, $f_{sub}$ is a WaveNet-like \cite{wavenet} structure consisting of a stack of depthwise separable dilated convolutions and skip connections. 

 \subsection{Conditioning network}
The baseline SE-Flow model uses conditional signals to guide the network to generate the target output. As conditional pre-processing the baseline model uses a single layer of depthwise separable convolutions, which are added to each subnetwork inside the coupling layer via a gated tanh activation. Therefore, conditional features are learned for each flow block individually, since there is no connection from one conditional injection to another.
In \cite{strauss_slt} it was already shown that a perceptually motivated pre-processing can guide the flow model to produce a better quality outcome.
Still, we assume that more effective conditioning requires high-level
representations, that are learned by making use of the representations
learned by previous steps. This representations can be particularly rich given the depth of such models, which usually consist of 12 or more blocks.

Because of this, we use a separate conditional network termed \textit{condNet}, which is implemented to create more meaningful representations from the conditional input. 
The condNet consists of two parts illustrated in Figure \ref{fig:arch}\,a). First, the encoder structure of \cite{macartney2018improved} is used with stride 1 to keep the spatial dimension which is needed for the affine transformation in the coupling layers. The output of each encoder layer is forwarded to an additional conditional block to match the feature dimension of the flow block. 

 \begin{figure}[htb]\captionsetup[subfigure]{font=large}
 \centering
\resizebox{0.43\textwidth}{!}{

\subfloat[]{\includegraphics[]{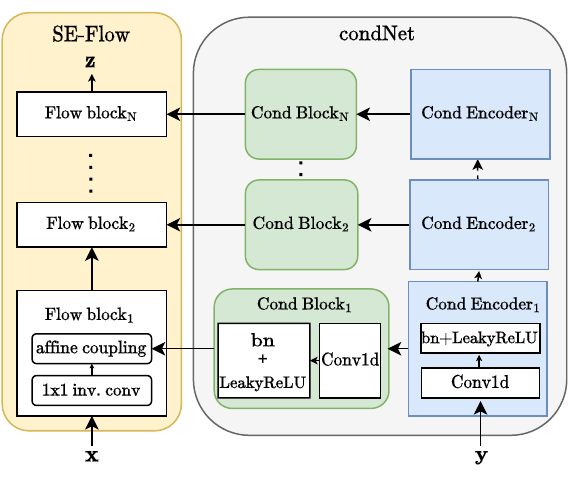}}\hspace{7mm} 
\subfloat[]{\includegraphics[]{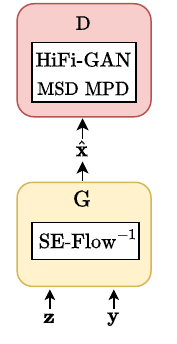}}
}
\caption{a) The architecture of SE-Flow using the condNet. The output of each encoder layer is used as input for the conditional blocks preparing the input for each coupling layer. b) The inverse SE-Flow serves as the generator (G) in the GAN training. G takes Gaussian noise $\textbf{z}$ and noisy speech $\textbf{y}$ to generate enhanced speech utternace $\hat{\textbf{x}}$. The discriminator (D) is the one proposed in \cite{hifigan} with multi-scale (MSD) and multi-period (MPD) discriminators.}
\label{fig:arch}
\end{figure}

\subsection{GAN training}
In initial experiments we found that using the condNet structure improves the overall quality, but a gap to the best performing models remains.
To bridge this gap we train the model further adversarially to guide it to produce more realistic samples.
Although GAN training often suffers from an unstable training process, we observe  stable training and fast convergence for our models. 
The overall generator training objective for SEFGAN is a commonly used combination of adversarial, feature matching (FM) and a reconstruction loss
\begin{equation}
    \mathcal{L}_{G} = \mathcal{L}_{adv}(G;D) + \mathcal{L}_{FM}(G;D) + \mathcal{L}_{rec}(G),    
\end{equation}

\noindent where $G(\textbf{z}, \textbf{y})=f_\theta^{-1}(\textbf{z}, \textbf{y})$ is the inverse flow.
We use the least-squares adversarial loss $\mathcal{L}_{adv}$ and the multi-resolution STFT recostruction loss $\mathcal{L}_{rec}$ defined in \cite{multires_stft}. In addition, we experimented with other reconstruction losses, e.g. Mel-Spectogram distance and SI-SDR loss, but we informally found the multi-resolution STFT to perform best.
The discriminator used is the one proposed in \cite{hifigan}.
It consists of an ensemble of $K=8$ discriminators working on different time scales and periods, leading to the total discriminator loss
\begin{equation}
    \mathcal{L}_{D} = \sum_{k=1}^K \mathcal{L}_{Adv}(D_k;G),
\end{equation}

\noindent where each $D_k$ is a different discriminator in the ensemble.

Evaluating the likelihood of SEFGAN indicates that the GAN training leads the model to disconnect from the maximum likelihood estimation (see Figure \ref{fig:hybrid}\,a)).
There are number of scenarios where we are interested in having a reliable log-likelihood estimation, e.g. for anomaly detection or confidence estimation. 
Moreover, it is a known effect of conditional GANs that they learn to ignore the latent noise input \cite{isola2017image}.

To overcome these issues we apply a hybrid training approach combining likelihood and adversarial training. For likelihood estimation, the NF model generates noise in the forward pass, while the adversarial training requires an audio sample generated from noise by the inverse model. For both steps a gradient update is performed.
The training objective in this case becomes 
\begin{equation}
    \mathcal{L}_{hybrid} = \mathcal{L}_{G} + \lambda \mathcal{L}_{nll},
\end{equation}
\noindent where $\lambda$ is a weighting factor of the likelihood training. 
\section{Experimental setup}
\label{sec:experiments}

\subsection{Dataset}
The dataset used for the experiments was proposed in the work of \cite{richter2023} and is a combination of the WSJ dataset \cite{wsj} and noise samples from CHiME3 \cite{chime3}. The signal to noise ratios (SNRs) are sampled uniformly between 0 and 20\,dB. The sampling rate was 16\,kHz. With this, a total duration of training and validation data of 24.92\,h using 101 different speakers was obtained. The test set includes 651 utterances (1.48\,h) from 8 different speakers in the same SNR range. 

\subsection{Model configuration}
\label{sec:configuration}
For the NF model we used 20 flow blocks and a subsampling factor of $s=12$. The subnetwork $f_{sub}$ had 8 layers of dilated convolutions with 128 output channels. As in \cite{Prenger2019}, the model uses early channel outputs to the loss function to create a multi-scale architecture. The condNet used an increasing amount of output channels by factor 24 before being mapped to 256 via the cond block. The kernel size was set to 15.

The flow models were trained using an early stopping mechanism if the validation loss did not decrease for 40 consecutive epochs. A batch size of 16 and an Adam optimizer with an initial learning rate of 0.001 was used for all models. The learning rate was reduced by a factor of 0.8 with a patience of 10 epochs.

For the GAN training all models were trained for 200 epochs using the Adam optimizer ($\left[\beta_1=0.5, \beta_2=0.9\right]$). The learning rate for the generator and discriminator was set to $5.0e-5$ and $2.0e-4$ respectively, together with an exponentially decreasing learning rate (factor: 0.8). The weighting factor in the hybrid training was experimentally selected to be $\lambda=0.3$. 

\subsection{Ablation study}

\begin{table}[t!]
	\centering
	\caption{Ablation study of the proposed system.}
 \resizebox{0.33\textwidth}{!}{
	\begin{tabular}{@{}lcc@{}}		
		\toprule
		\textbf{Model}                   & \textbf{SI-SDR [dB]} & \textbf{2f-model} \\ \midrule
		Baseline (SE-Flow \cite{strauss2021})      &    13.94   &  36.58 \\
		w. condNet                &   16.50   & 47.59   \\
		w. reconstruction loss &   18.82   & 50.51   \\
		SEFGAN              &  18.76   &  52.44   \\ \bottomrule
	\end{tabular}}
 \label{tab:ablation}
\end{table}

We performed an ablation study based on SI-SDR \cite{sisdr} and 2f-model score \cite{2f} in order to investigate the effect of the different components of SEFGAN. These metrics were selected to give us an indication about introduced distortions and perceptual quality of the enhanced samples. The results are shown in Table \ref{tab:ablation}. 
It can be seen that all components add to the quality of SEFGAN, with adding the condNet giving the strongest initial improvement of 3\,dB in SI-SDR and 11 points in 2f-model score. This can be improved by adding a reconstruction loss on the inverted network and train in a discriminative way. Using the SEFGAN results in a slightly reduced SI-SDR \cite{sisdr} value compared to only adding a reconstruction loss on top, but a higher 2f-model score suggesting that the model is able to create better sounding results.

\subsection{Comparing methods}
The proposed \textbf{SEFGAN} was compared to a set of competitive models. As a baseline, we retrained the \textbf{SE-Flow} model from \cite{strauss2021} using our configuration without and with the condNet (\textbf{SE-Flow+condNet}). This model was then utilized as generator in the SEFGAN framework. The hybrid learning approach is labeled \textbf{SEFGAN-hybrid}. \textbf{SGMSE+} \cite{richter2023} is a state-of-the-art DDPM for speech enhancement. To generate the enhanced samples we used the checkpoint provided by the authors, which was trained on the same dataset.
\textbf{MetricGAN+} \cite{fu21_interspeech} is a GAN-based masking system optimized directly on PESQ. We used the SpeechBrain \cite{speechbrain} implementation and setting for comparison. \textbf{Conv-TasNet} \cite{Luo2019ConvTasNet} is a popular mask-based discriminative system with learned encoder-decoder structure, initially developed for speaker-speaker separation. MetricGAN+ and Conv-TasNet were retrained on the data. 

\subsection{Evaluation}
\textbf{Computational metrics:} A set of computational metrics was selected to evaluate the proposed system in comparison to similar methods. We use PESQ \cite{pesq} (worst:\,-0.5; best:\,4.5), eSTOI \cite{estoi} (worst:\,0; best:\,1), SI-SDR (in dB) and 2f-model score (worst:\,0; best:\,100). 
We also report the word error rate (WER) of the samples from the lower third of the SNR range of the test set. This is  relevant for intelligibility as well as for a potential application of the models as a front-end to a speech recognition system. We used the transcriptions included into the CHiME challenge \cite{chime3} and an open source speech recognizer from SpeechBrain \cite{speechbrain}.

\noindent \textbf{Listening test:} A listening test based on the MUSHRA methodology \cite{mushra2015} was conducted to evaluate the perceptual performance. Therefore, one random sample per speaker from the lower third of input SNRs in the test set was selected, resulting in 8 test items. For a fair comparison between the systems we performed background matching as described in \cite{strauss_slt} with a background attenuation in the reference condition of 35\,dB. 
The test included a low pass anchor with 3.5\,kHz cutoff frequency (lp35) as low quality orientation point. 
We used the webMUSHRA platform \cite{webmushra} with 13 expert listeners as participants on their own laptop and pair of high quality headphones.
The participants were asked to rate the overall audio quality of the presented items. Computational metrics evaluation was repeated on the selected items confirming the same overall trend as in the computational metrics on the full test set. The test items can be found online: \url{https://www.audiolabs-erlangen.de/resources/2023-WASPAA-SEFGAN}. 

\section{Results and discussion}
\label{sec:results}

\subsection{Listening test results}
The results of the listening test can be found in Figure \ref{fig:lt_res}. They confirm the strong improvement in perceptual outcome of using SEFGAN in comparison to SE-Flow\,+\,condNet with a difference of 15 MUSHRA points. SEFGANs results lie in the \textit{good} to \textit{excellent} area, just behind SGMSE+.
Conv-TasNet takes the third place behind SEFGAN still in the \textit{good} quality range. On average, SGMSE+ shows the best perceptual performance among the tested methods lying in the \textit{good} and \textit{excellent} range. 
MetricGAN+ demonstrates the worst performance with an average result below the low-pass anchor. This is in accordance with the results in \cite{strauss_slt} indicating that MetricGAN+ struggles in low SNR ranges.\\
It is worth pointing out that the perceptual evaluation was only conducted for the most difficult SNR settings, since we noticed in an informal evaluation that for the higher SNR conditions all models perform comparably well.

\begin{figure}[t!]
	\resizebox{0.47\textwidth}{!}{
		\centering
		\begin{minipage}[b]{\textwidth}
			\includegraphics[width=1\textwidth]{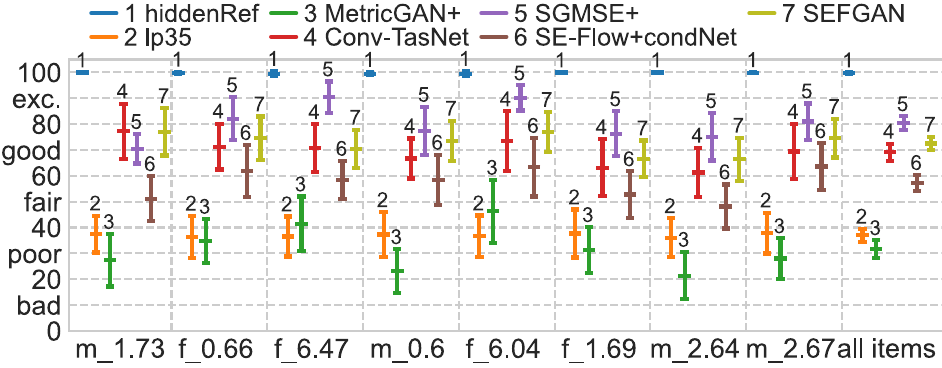}
	\end{minipage}}
 \caption{Listening test results. Student's t-distribution with mean and 95\% confidence intervals. The item names indicate male (m) or female (f) speakers, as well as, input SNR (in dB) condition.}
 \label{fig:lt_res}
 \end{figure}

\subsection{Computational evaluation results}


\begin{table}[]
\centering
\setlength{\tabcolsep}{3.2pt}
\caption{Computational evaluation results on the WSJ-CHiME3 dataset (Mean; Best results in bold).}
\resizebox{0.44\textwidth}{!}{
	\begin{tabular}{@{}l|cccc|c@{}} 
		\toprule
	       & \textbf{PESQ$\uparrow$} & \textbf{eSTOI$\uparrow$} & \textbf{SI-SDR$\uparrow$} & \textbf{2f-model$\uparrow$}& \textbf{WER $\downarrow$}\\ \midrule %
		Conv-TasNet \cite{Luo2019ConvTasNet}        &  2.91     &   \textbf{0.93 }    &    \textbf{19.69}   &  51.75  & 10.75     \\
		MetricGAN+ \cite{fu21_interspeech}         &  2.90   &  0.88     &  12.14      &   36.68  &  14.45 \\ 
		SGMSE+  \cite{richter2023}           &   2.91   & 0.92       &  17.88     &  48.39  & 12.24    \\ 
	SE-Flow \cite{strauss2021} & 2.12  & 0.83  & 13.94  & 36.58 & 35.29  \\ \midrule \midrule
 SE-Flow + condNet    &  2.53    &   0.89   &     16.50           &   47.59  & 16.20     \\
        SEFGAN-hybrid  &   2.91    &   \textbf{0.93}   &   17.65           &   50.39  & \textbf{10.60} \\
    SEFGAN  &   \textbf{2.94}    &   \textbf{0.93}   &   18.76   &           \textbf{52.44}  & 10.96\\ \bottomrule
\end{tabular}}
\label{tab:results}
\end{table}
The results of the computational evaluation are shown in Table \ref{tab:results}. As indicated by the ablation study, SEFGAN outperforms the baseline SE-Flow model with condNet in all metrics confirming the benefits of the proposed system. The metrics also show that SEFGAN, performs best among all models in PESQ, eSTOI and 2f-model. Conv-TasNet is on par with SEFGAN in eSTOI and shows the strongest results in SI-SDR. This can be explained by the fact that the model is directly optimized on that metric. SE-Flow has the lowest values in all metrics followed by MetricGAN+. SGMSE+ shows higher values in all metrics than SE-Flow\,+\,condNet  and SE-Flow, but is behind SEFGAN and Conv-TasNet. \\ \indent
Analysing the WER it can be seen that SEFGAN-hybrid performs best, slightly before SEFGAN and Conv-TasNet. A closer inspection of the samples generated via SGMSE+ reveals a strong denoising performance, but also occasionally adding breathing-like noise sounds and a loss of word parts. This could explain the objective metrics evaluations especially for the WER. The WER results suggest that Conv-TasNet and SEFGAN are a more robust choice in a scenario as a pre-processor for a speech recognition system.

\subsection{Analysis of computational complexity}
\begin{table}[htb]
\centering
	\setlength{\tabcolsep}{3.2pt}
	\caption{Parameter and real-time factor comparison.}
 \resizebox{0.24\textwidth}{!}{
\begin{tabular}{@{}lcc@{}}
\toprule
\textbf{}            & \textbf{\#Params} & \textbf{RTF} \\ \midrule
Conv-TasNet &   \phantom{0}3.5M      & 0.20  \\
SGMSE+      &  65.6M       &   6.28  \\
SE-Flow & 11.0M & 0.35 \\
SE-Flow\,+\,condNet & 34.8M & 0.38\\
SEFGAN      &  34.8M      &  0.38  \\ \bottomrule
\end{tabular}}
\label{tab:complex}
\end{table}
\noindent Table \ref{tab:complex} shows a comparison of computational complexity in terms of number of parameters and real-time factor (RTF) between Conv-TasNet, SGMSE+, SE-Flow, SE-Flow\,+\,condNet and SEFGAN. The reported numbers are the average processing time for 10 audio files on a machine with an NVIDIA Quatro RTX 4000 GPU and an Intel Core i7-10850H CPU @ 2.70GHz. 
The results show that despite adding more parameters to the enhancement network, SE-Flow\,+\,condNet and SEFGAN only have a marginally increased RTF compared to SE-Flow. The DDPM has the largest complexity in both amount of parameters and RTF by a large margin. The large RTF is expected as diffusion models require multiple steps to complete one inference step, while SEFGAN, the NF-based models and Conv-TasNet generate their samples in one forward pass.

\subsection{Likelihood analysis}
\begin{figure}[htb]\captionsetup[subfigure]{font=large}
\centering
\resizebox{0.45\textwidth}{!}{
\subfloat[]{\includegraphics[height=3in]{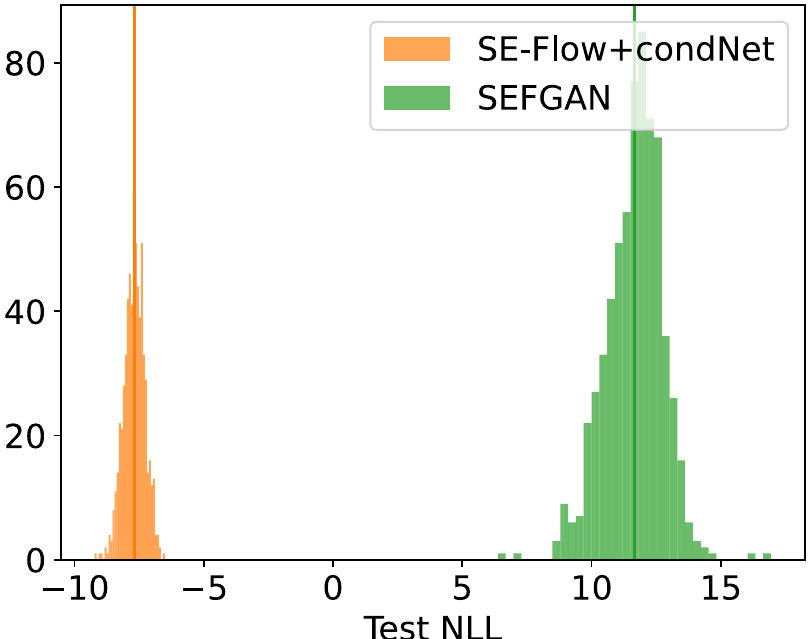}}
\subfloat[]{\includegraphics[height=3in]{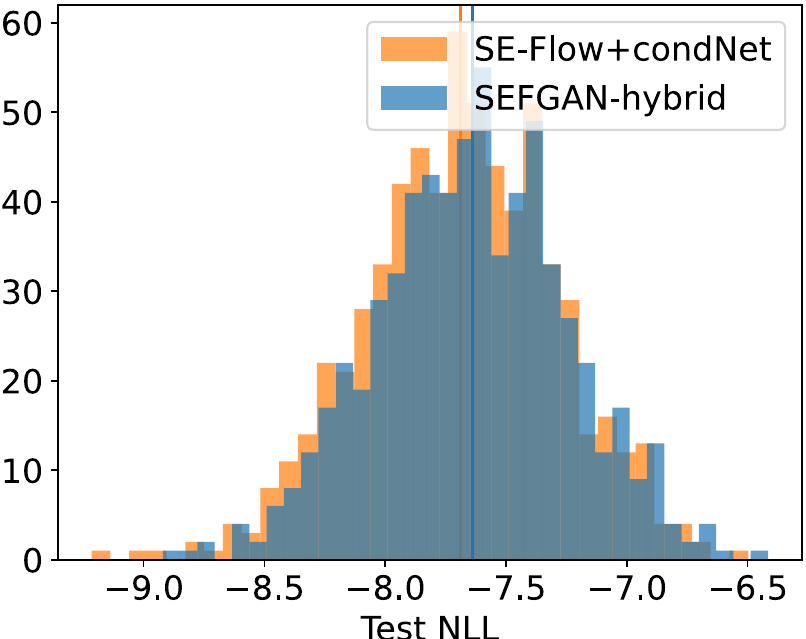}}}
\caption{The distribution of test likelihood for the SE-Flow\,+\,condNet system in comparison to the SEFGAN and a SEFGAN-hybrid approach. The hybrid approach matches the likelihood to the original distribution.}
\label{fig:hybrid}
\end{figure}

 \noindent Figure \ref{fig:hybrid} shows the effect of the hybrid training approach where the test NLLs match the SE-Flow\,+\,condNet results. The findings are also in accordance with \cite{flowgan}. In Table \ref{tab:results} the computational evaluation metrics are reported for the hybrid approach showing that this method still reaches comparable results to the other models. Inspecting a few generated samples indicates that a slightly worse perceptual result would be obtained, suggesting a trade-off between pure SEFGAN and SEFGAN-hybrid approach depending on the application. 
 Further investigation is left for future work. Still, it is worth highlighting that this result gives SEFGAN-hybrid an advantage over discriminative models like Conv-TasNet in applications where likelihood estimation is needed.
 
\section{Conclusion}
\label{sec:conclusion}
In this paper a combination of NF-based models and GANs for speech enhancement called SEFGAN was proposed. In particular, the trained flow model was used as a generator to denoise corrupted speech samples. It was shown that this approach substantially improves the pure NF-based technique, without adding further complexity in inference. The model produces high-quality speech as measured by computational evaluation metrics and a listening test. Further analysis shows that a simple hybrid learning approach allows the model to keep its maximum likelihood estimation capabilities while still improving the quality of the estimated speech signals. Finally, the model functions well in tandem with an ASR model making it suitable for a wide range of applications.\newline
In the future, we plan to investigate the potential of combining our approach with a MetricGAN discriminator. 
\bibliographystyle{IEEEtran}
\bibliography{refs23}

\end{sloppy}
\end{document}